\documentclass[a4paper,11pt]{article}

\usepackage{jheppub} 
\usepackage[T1]{fontenc} 
\def\beq{\begin{equation}}
\def\eeq{\end{equation}}

\newcommand{\bea}{\begin{eqnarray}}
\newcommand{\eea}{\end{eqnarray}}

\title{\boldmath CFT duals of Kerr-Taub-NUT and beyond.}

\author[a,b,c]{Malcolm J. Perry,\ }
\author[d,e]{Maria J. Rodriguez\ }

\affiliation[a]{Department of Physics and Astronomy, Queen Mary University of London, Mile End Road, London E1 4NS, UK.}
\affiliation[b]{DAMTP, Centre for Mathematical Sciences, Wilberforce Road, Cambridge, CB3 0WA, UK.}
\affiliation[c]{Trinity College, Cambridge, CB2 1TQ, UK.}
\affiliation[d]{Department of Physics, Utah State University, 4415 Old Main Hill Road, UT 84322, USA.}
\affiliation[e]{Instituto de Fisica Teorica UAM/CSIC, Universidad Autonoma de Madrid,13-15 Calle Nicolas Cabrera, 28049 Madrid, Spain.}
\emailAdd{malcolm@maths.cam.ac.uk}
\emailAdd{maria.rodriguez@gmail.com}

\abstract{The duality relating  the four-dimensional Kerr-Taub-NUT black hole to a thermal two-dimensional CFT
with central charges $c_L=c_R=12 J_0$ is analyzed in detail, generalizing an argument given recently for Kerr within the soft-hair approach. The
hidden conformal symmetry is realized in the form of $Vir_L \times Vir_R$ diffeomorphisms which act non-trivially on the black hole horizon.
Semiclassical formulae are derived for the temperature and central charges of the dual CFT.
Assuming the applicability of the Cardy formula, these CFT quantities precisely reproduce the macroscopic Bekenstein-Hawking area law.
 Various further generalizations including the complete family of black holes in four dimensions are discussed.}

\begin{document} 
\maketitle
\flushbottom

\section{Introduction}

Many black holes in 4-spacetime dimensions, including Kerr black holes with spin J, are conjectured to have a dual two-dimensional conformal field theory (2D CFT) description with central charges $c_L=c_R=12\, J$. If one assumes the applicability of the Cardy formula, the central charges reproduce the macroscopic area-entropy law. The synthesis between 4D-gravity and 2D CFT can be obtained in part from the fact that we can relate the description of the
four-dimensional black hole horizon as a quotient of a deformed AdS$_3$ to the conformal field theory description
\footnote{More precisely, the black hole near bifurcation surface metrics are quotients of the deformed AdS$_3$ geometries and a `$\theta$-leave' over the bundle.}. This turns out to be enough information to show that the extremal 4D-black holes are dual to two-dimensional CFT \cite{Guica:2008mu,Hartman:2008pb}.

Another fascinating result stemming from the symmetry analysis of the black hole Klein-Gordon equation (rather than an analysis of the spacetime) led to the observation that generic non-extreme black holes are dual to finite 2D CFT at finite temperatures \cite{Castro:2010fd}. The ‘hidden conformal symmetry’ which act on solutions of the scalar wave equation at low frequencies seems to persist for generic non-extreme values of the black hole spin $J$.

A more refined version of the duality between gravity in 4D and a CFT comes from the so-called soft hair approach. It was recently shown in \cite{Haco:2018ske} that the hidden conformal symmetry in the form of $Vir_L \times Vir_R$ diffeomorphisms act non-trivially on the black hole horizon generating soft hair. This soft hair is realized as finite covariant right(left)-moving Iyer-Wald Virasoro charges and the central terms.
To guarantee the integrability and associativity of the charges a Wald-Zoupas counterterm was proposed. Despite these developments there has been relatively little effort to prove uniqueness of the counterterm. Assuming the validity of the Cardy formula, it is now clear that with the Wald-Zoupas counterterm originally proposed in \cite{Haco:2018ske} only reproduces the area law for the Kerr(Newman) black hole \cite{Haco:2018ske,Haco:2019ggi}. Already in \cite{Perry:2020ndy}, the authors found that for AdS-Kerr black holes an alternative proposal to the Wald-Zoupas counterterm which was consistent with all previous works \cite{Hartman:2008pb,Lu:2008jk}. Further, this generalized term produced the dual 2D CFT at temperatures $(T_L,T_R)$  and central charges consistent with the thermodynamics of these black holes.

Previous potentially related attempts to validate the Wald-Zoupas counterterm  for 4D black hole solutions include \cite{Chen:2020nyh,Chandrasekaran:2020wwn}.

In this paper we take steps towards a more global understanding of the counterterm uniqueness by studying a generalized version of the counterterm valid for the complete family of black holes in 4D. The linearized charge contribution from the proposed counterterm to the charges is defined by
\bea\label{eq:counterterm}
\delta \mathcal{Q}_{ct} =\frac{1}{16\pi}\int_{\partial\Sigma} {F_{(ct)}}_{ab}\, d \Sigma^{ab}\,
\eea
where $N$ is the volume two-form on the normal bundle to the $\Sigma_{bif}$. 
\bea
{F_{(ct)}}_{ab}=- 4 {N_{d}}^c\nabla_c(\zeta_{[a} {h^d}_{b]}) \,,
\eea
where $\zeta$ is associated to a diffeomorphisms acting on the horizon, and $h$ an on-shell linearized fluctuation around a fixed black hole background.

Our most interesting results concerns the soft hair charges of the Kerr-Taub-NUT black hole spacetimes  with spin $J_0$. We find that the counterterm (\ref{eq:counterterm}) satisfies the known consistency requirements and yields central charges $c_L=c_R=12 \, J_0$. One would like to go beyond this and explicitly identify the 2D CFT description within the complete family of Plebanski-Demianski black hole classes. In all cases, employing the generalized counterterm (\ref{eq:counterterm}), within the soft hair and monodromy approach we are able to determine the central charges $c_L=c_R$ and dual temperatures. We then apply the thermodynamic Cardy formula relating the microscopic entropy of a unitary CFT to its temperature and central charge. The resulting entropy agrees exactly with the area-entropy law, providing corroboration for our proposal that generic black holes in 4D-gravity are dual to a two-dimensional CFT. 

We can list the proposed central charges $c_L=c_R$ and dual temperatures in 4D-gravity/ 2D CFT be extreme or not. Table \ref{Table1} summarizes the results obtained using Kerr/CFT, soft hair approach, monodromy technique during the past years. For the details on this published work we refer the reader to the papers cited \footnote{The temperatures and central charges for Kerr-Taub-NUT are 
derived in the present paper.}.

\begin{table}[!]
\centering
\begin{tabular}{|c | c | c | c |}
\hline
Black hole &  Central charge & CFT Temperatures  & Reference  \\
\hline
\hline
extreme Kerr & $c_L=12\,J$ &  $ \begin{array} {lcl} T_R&=&  0 \\  T_L&=&\frac{1}{2\pi} \end{array} $  & \cite{Guica:2008mu}  \\
\hline
Kerr & $c_R=c_L=12\,J$ & $ \begin{array} {lcl} T_R&=&  \frac{r_+-r_-}{4\pi a}\\ T_L&=&\frac{r_++r_-}{4\pi a}\end{array} $  &   \cite{Castro:2010fd} \cite{Haco:2018ske} \cite{Castro:2013kea} \\
\hline
extreme Kerr-Newman &  $c_L= \frac{24 a r_+}{\Delta''(r_+)}$ & $ \begin{array} {lcl} T_R&=&  0 \\  T_L&=& \frac{2 r_+^2-q^2}{4\pi r_+\sqrt{r_+^2-q^2}} \end{array} $   & \cite{Hartman:2008pb}\\
\hline
Kerr-Newman &$c_R=c_L=12\,J$  &  $ \begin{array} {lcl} T_R&=&  \frac{r_+-r_- }{4\pi a}\\ T_L&=&\frac{(r_++r_-)^2 - 2 Q^2}{4\pi a (r_++r_-)}\end{array} $  & \cite{Haco:2019ggi} \cite{Wang:2010qv} \cite{Chen:2010xu} \\
\hline
extreme Kerr-AdS&  $c_L = \frac{24 a r_0}{\Delta''(r_0)}  $ & $ \begin{array} {lcl} T_R&=&  0 \\  T_L&=&\frac{(r_0^2+a^2)\Delta''(r_0)}{8\pi r_0 \Xi} \end{array} $ & \cite{Hartman:2008pb} \cite{Lu:2008jk} \\
\hline
Kerr-AdS & $c_R=c_L=\frac{ 6 a (r_+^2-r_-^2)}{\Delta'(r_+)}$ & $ \begin{array} {lcl} T_R&=&\frac{\Delta'(r_+)}{4\pi \,a\Xi}\\ T_L&=&\frac{(\Omega_++\Omega_-)}{(\Omega_- -\Omega_+)}\frac{\Delta'(r_+)}{4\pi \,a\Xi }\end{array} $\,
  &  \cite{Perry:2020ndy} \\
 \hline
extreme Kerr-Taub-NUT& $c_L=12 a M $&  $ \begin{array} {lcl} T_R&=& 0\\ T_L&=&\frac{r_+^2 + N^2}{2\pi a r_+}\end{array} $ &  \cite{Setare:2010cj} \\
 \hline
Kerr-Taub-NUT & $ c_R=c_L=12\,J_0 $&  $ \begin{array} {lcl} T_R&=& \frac{r_+-r_-}{4\pi a}\\ T_L&=&\frac{(r_++r_-)^2+4 N^2}{4\pi a (r_++r_-)}\end{array} $ & 
\\
\hline
\end{tabular}
\caption{The central charge $(c_L,c_R)$ and 2D CFT dual (left)right-temperatures $(T_L,T_R)$ for 4D black holes that reproduce the macroscopic Bekenstein-Hawking area law assuming Cardy formula $S=\frac{\pi^2}{3}(c_R \,T_R+c_L \,T_L)$. All entries share the same central charge pattern; the right/left central charge in the  2D CFT description for 4D black holes is suggested to be equal in each sector $c_L=c_R$.  Extreme black holes are defined as stationary black holes with vanishing Hawking temperature $T=0$ when the outer and inner horizons are set equal to each other $r_+=r_-$. This limit identifies also the dual temperature $T_R=0$ and thus predicts no contribution to the entropy from this sector. As a result, it is not possible to determine the central charge values $c_R$ for extremal black holes. This limit contrasts with the prediction for non-extremal black holes where $T_R \ne 0$ where $c_R$ can naturally be accounted.}
\label{Table1}
\end{table}

This paper is organized as follows. In section \ref{sec:KTN}, we will review the Kerr-Taub-NUT black hole metric. We will also
present the Smarr law and first law for the solution. In section \ref{sec:MonANdConfCoord}, we will briefly discuss the so-called monodromy technique \cite{Castro:2013kea} that also serves to identify the conjugate variables, and dual 2D CFT quantities. The conformal coordinates in which the Virasoro action takes the simple form are presented in section \ref{sec:ConfCoord}. The linearized covariant charges are also found in this section and the central terms computed. The identifications we previously derived for the (T$_L$,T$_R$) temperatures are also shown to arise from imposing the first law of thermodynamics in section \ref{sec:FirstLaw}.  In Section \ref{sec:Misner} we take the rotating Taub-NUT black hole solution with an overall distribution of the Misner string, and show that similar identifications yield a 2D CFT dual.
Further generalizations including the computation of a well-defined charge for the most general 4D black hole solution of General Relativity are discussed in Section
\ref{sec:AllBH}. In the discussion section we present some final remarks. Further technical points are relegated to the appendices.\\
Throughout this paper we use units such that $c=k=G=\hbar=1$.

\section{Kerr-Taub-NUT}
\label{sec:KTN}

In this section we review the Kerr-Taub-NUT black hole and fix our notation. This will provide the setup for studying the covariant charges and computing the central terms in Section \ref{sec:ConfCoord}. 

We begin by considering the four-dimensional Kerr-Taub-NUT black hole solution which has also been recently the center of many other works \cite{Crawley:2021auj,Rodriguez:2021hks}.
The Kerr-Taub-NUT metric in Boyer-Lindquist co-ordinates $(t,r,\theta,\phi)$  is
\beq\label{KerrTaubNUT}
ds^2 = \rho^2 \left(\frac{dr^2}{\Delta}+d\theta^2\right)+\frac{\sin^2\theta}{\rho^2}(-a \,dt+\Sigma \,d\phi)^2 - \frac{\Delta}{\rho^2}(dt-\Xi \, d\phi)^2
\eeq
where
\bea\label{eq:functions}
&&\rho^2=r^2+(N+a \cos\theta)^2\,,\,\,\, \Delta=r^2-2 m r-N^2+a^2\,,\\
&& \Xi=a\sin^2\theta - 2N \cos\theta\, ,\,\,\,\,\,\, \Sigma=r^2+N^2+a^2\,
\eea
This solution to Einstein's vacuum equations $ R_{\mu\nu} - (1/2)\, R\, g_{\mu\nu}=0$ carries three parameters: $m, a , N $ which represent respectively the mass, rotation and nut charge. The location of the outer/inner horizons defined via $\Delta=0$ are $r_{\pm}=M\pm\sqrt{M^2-a^2+N^2}$.

With these geometric quantities in hand, let us now turn towards thermodynamics. The corresponding thermodynamic quantities can be found in \cite{Bordo:2019rhu}  \footnote{ The Kerr-Taub-NUT solution considered in our paper corresponds to $s= 0$ solutions in \cite{Bordo:2019rhu} containing two Misner strings that are ‘symmetrically distributed’ making both axes equally singular.}. The Kerr-Taub-NUT black hole exhibits a mass, angular momentum and angular velocity defined by
\bea
M=m\,,\qquad J= \frac{a \Sigma_+ }{2r_+}=a\left(M+\frac{N^2}{r_+}\right)\,,\qquad \Omega_{\pm}=\frac{a}{\Sigma_{\pm}}\,.
\eea
The entropy and temperature can be assigned as
\bea
S_{\pm}=\pi \,\Sigma_{\pm}\,,\qquad T_{\pm}=\frac{\kappa_{\pm}}{2\pi}=\frac{ \Delta'(r_{\pm})}{4\pi \Sigma_{\pm}}\,,
\eea
and NUT charge and Misner potential by
\bea
N_{\pm}=-\frac{2\pi N^2 (N\mp a)}{r_+}\,,\qquad \psi_{\pm}=\frac{1}{8\pi N}\,.
\eea
where $\Sigma_{\pm}=\Sigma(r_{\pm})=r_{\pm}^2+a^2+N^2$ and $ \Delta'(r_{\pm})=\pm(r_+-r_-)$. These thermodynamic quantities obey a generalized first law:
\bea\label{first law}
\delta M=T\delta S +\Omega\, \delta J+\psi_+ \delta N_+ +\psi_- \delta N_-\,,
\eea
together with the corresponding Smarr relation
\bea
M=2(T S+\Omega \, J+ \Psi_+N_++\Psi_-N_-)\,.
\eea

Let us also note the following two interesting facts about the total angular momentum $J$.  The total angular momentum $J$ actually differs from the asymptotic charge $J_0= a M $ by the Misner string contribution $J_s=J-J_0= a N^2/r_+$. Both $J_s$ and $J_0$ are also in agreement with the values directly computed via the Komar integrals.  As we will now show the asymptotic charge $J_0$ will become the relevant quantity for the CFT description.

\section{Monodromies and 2D CFT}
\label{sec:MonANdConfCoord}

The monodromy technique developed in \cite{Castro:2013kea} relies on the monodromy data of the Klein-Gordon equation in a curved background to identify the (T$_L$,T$_R$) temperatures of the dual 2D CFT. As we now describe, using the monodromy technique we are able to both successfully apply the method to a larger body of data and to derive the conjugate 2D CFT values for the Kerr-Taub-NUT black hole solution. In this section we also describe, assuming the Cardy formula, the corresponding  central charge for this solution.

The focus in this section is the Klein-Gordon equation for a scalar massless field 
\bea
 \frac{1}{\sqrt{-g}}  \partial_{\mu}(\sqrt{-g}\,g^{\mu\nu}\,\partial_{\nu}\Psi)=0\,.
 \eea 
 One can easily verify the equation is separable in the Kerr-Taub-NUT background (\ref{KerrTaubNUT}). The radial equation for $\Psi(t,\phi,r,\theta)=e^{-i\omega t+im\phi}R(r)S(\theta)$ reduces to 
\beq\label{eq:Req}
\partial_r\left[\Delta\partial_r R(r)\right]+\left[K_l+\frac{(\Sigma \,\omega- a m)^2}{\Delta}\right]R(r)=0\,,
\eeq
Here $K_l$ is the separation constant and the functions $\Delta,\Sigma$ are defined in (\ref{eq:functions}).
This equation contains three singularities: two regular singular points located at the outer/inner horizons $r=r_{\pm}$ and an irregular singular point at $r=\infty$. Solutions to (\ref{eq:Req}) will have branch cuts at the regular singular points. One can immediately see that the series expansion of the two linearly independent solutions around $r=r_{\pm}$ points can be expanded respectively as
\bea\label{solutions}
R(r)=(r-r_+)^{\pm i\,\alpha_{+}} [1+ O(r-r_+)]\qquad  \text{and} \qquad R(r)=(r-r_-)^{\pm i\, \alpha_{-}} [1+ O(r-r_-)]\,.
\eea

 The monodromies at each of these regular singular points are:
\bea\label{monodromies}
\alpha_{\pm}= \frac{(\omega-\Omega_{\pm}m)}{2\,(\pm \kappa_{\pm})}\,.
\eea
This monodromy data possesses highly intricate but universal properties which follow from geometry alone. Moreover, recent analyses indicate that the monodromy data encodes information about black hole thermodynamics, hidden conformal symmetry and evidence for a 2D CFT description of the thermal properties of black hole microstate \cite{Chanson:2020hly,Aggarwal:2019iay,Puletti:2021msz}.

Following \cite{Castro:2013kea} we consider instead the linear combination of the monodromies \footnote{Note that the monodromy definitions here $\alpha_{\pm}$ are related to the definitions in \cite{Novaes:2014lha} via $\pm i 2 \alpha_{\pm}=\theta_{\pm}$.} for the energy eigenvalues $\omega_{L,R}$ of $(i\partial_{t_R},i\partial_{t_L})$
\bea
\omega_L=\alpha_+-\alpha_-\,,\qquad
\omega_R=\alpha_+ + \alpha_-\,.
\eea
with  $t_{L,R}$  precisely  the  coordinates  in  the  monodromy  basis. Thus, we can write
\bea
\Psi(t,\phi,r,\theta)=e^{-i\omega t+im\phi}R(r)S(\theta)=e^{+i \omega_L \,t_L-i \omega_R \,t_R} R(r)S(\theta)
\eea
using the explicit form of the monodromies (\ref{monodromies}) we find the conjugate variables \footnote{ In references \cite{Castro:2010fd} the conjugate variable are defined at $t_{\pm}$ such that the relation $e^{-i \omega t+i m \phi}=e^{-i \omega_L \,t_- -i \omega_R \,t_+}$. These are in agreement with our notation in that  $t_-=-t_L$ and  $t_+=+t_R$.}
\bea\label{eq:coord}
t_L=2\pi T_L \,\phi- \frac{1}{2M}\,t,\qquad
t_R= 2\pi T_R \, \phi\,,
\eea
with $(t,\phi)$ the Boyer-Lindquist coordinates and right/left temperatures 
\bea\label{temperatures} 
T_R=\frac{r_+-r_-}{4\pi a}\,,\qquad T_L=\frac{M^2+N^2 }{2\pi a M}=\frac{(r_++r_-)^2+4 N^2}{4\pi a (r_++r_-)}\,.
\eea
For vanishing  NUT charge $N=0$ we recover the Kerr black hole results \cite{Haco:2018ske}. 
Another interesting case is the extreme limit (see Table \ref{Table1}) for which $T_R\rightarrow 0$ as $r_+=r_-$.

Having found the conjugate CFT temperatures for the Kerr-Taub-NUT black hole, we would like to understand whether or not there is some way for us to recover the Bekenstein-Hawking formula for black hole entropy. Employing Cardy formula for a chiral CFT, we find the entropy obeys
\bea
S_{\pm}= S_{Cardy}=\frac{\pi^2}{3} c \, (T_L\pm T_R)\,,
\eea
provided the central charge $c=c_L=c_R$ for the black hole geometry obeys
\bea\label{centralcharge}
c=12 \,a \, M= 12\, J_0\,.
\eea
The CFTs have many universal properties that one would like to map. So far in our explicit matching between gravity and the CFT, we have found agreement between the Cardy and the Bekenstein - Hawking entropy formulas. Rather than trying to prove  that the Cardy formula is applicable, we take the precise entropy agreement resulting from our monodromy analysis as the first piece of evidence of this match. We bring this numerological observation a few steps closer to an actual explanation of the entropy. Within the soft-hair approach \cite{Haco:2018ske} we now turn to the vector fields which generate (local) symmetries of the near-bifurcation surface Kerr-Taub-NUT geometry and compute the non-vanishing boundary charges. More evidence will appear below in section \ref{sec:FirstLaw}.

\section{Conformal coordinates and Covariant charges}
\label{sec:ConfCoord}
%
%
%
%
%
%
%
%
In this section we derive the relation $c_R= 12 J_0$ for the central charge of Kerr-NUT black holes. To this end, following the construction of linearized covariant charges associated to the diffeomorphisms acting on the horizon \cite{Haco:2018ske} we evaluate the central term in the Virasoro charge algebra.

First, we must define conformal coordinates that are well-adapted to an analysis of the 4D black hole mirroring that of the 3D BTZ black holes \cite{Maldacena:1998bw}. The richness follows in part from the fact that 4D black hole geometries contain locally on the horizon a three-dimensional metric that is a quotient of a deformed AdS$_3$. This resonates the duality between conformal field theory and quantum gravity on AdS$_3$, as well as the central charge of the conformal field theory, derived
some time ago in \cite{Brown:1986nw}. To make this manifest for the Kerr-Taub-NUT black holes (\ref{KerrTaubNUT}) we can now conjecture the conformal coordinates $(w^\pm,y)$ which seem to most clearly exhibit the conformal structure as follows
\bea\label{eq:CFT}
w^+&=&R(r)\,e^{t_R}\,,\nonumber\\
w^- &=&R(r)\,e^{t_L}\,\\
y&=&Q(r)\,e^{(t_L+t_R)/2}\,\nonumber
\eea
with $(t_L, t_R)$ defined in (\ref{eq:coord}). Applying the coordinate transformation (\ref{eq:CFT}) with $R^2+Q^2=1$ and the functions $R^2(r) = (r- r_+)/(r-r_-)$ we find that the black hole metric around the bifurcation surface $w^{\pm}=0$ (to leading and sub-leading order) becomes
\bea\label{eq:metricBif}
ds^2&=& \frac{4\,\rho_+^2}{y^2}dw^+ dw^-+\frac{16 M^2 a^2 \sin^2\theta }{y^2 \rho_+^2}dy^2+\rho_+^2 \, d\theta^2\nonumber \\
&-&\frac{8 M \, w^{+} }{ y^3 \,\rho_+^2}  (r_+-r_-) \left(\Sigma_+ + 2 N a \cos\theta  \right) dw^{-} dy\\
&+& \frac{64 M^2 a^2 \sin^2\theta \, w^{-}}{ y^3 } \left[\frac{(r_+ +M)} { 4 M \rho_+^2}-\frac{(4\pi M T_L-  \Xi)(4 \pi M (T_L+T_R)- \Xi)  }{8 M^2 \rho_+^2\sin^2\theta} \right] dw^{+} dy\nonumber \\
&+&...\,,\nonumber
\eea
where 
\bea
\rho^2_+\equiv \rho(r_+)^2=r_+^2+(N+a \cos\theta)^2\,,\qquad \Sigma_+ =r_+^2+N^2+a^2\,.
\eea
Interestingly,  this limit is finite only for the specific choice of (\ref{temperatures}). Moreover, this approach implies a very specific form of the geometry on the black hole bifurcation surface at leading order (\ref{eq:metricBif}); for Kerr-Taub-NUT the `$\theta$-leaves' of fixed polar angle take the form of a quotient of a deformed $AdS_3$.

For future reference, a quantity that will be later employed in the computations of the surface integrals involves the volume element at leading order on the bifurcation surface
\bea
\epsilon_{+- y \theta} =\frac{8 a M \sin\theta  \rho_+^2 }{ y^3}+ ... \,.
\eea
as well as the inverse metric
\begin{eqnarray}
&& g^{yy}\sim \frac{ y^2  \rho_+^2}{16 M^2 a^2 \sin^2\theta }\,,\qquad
 g^{\theta\theta}\sim\frac{1}{\rho_+^2}\,,\qquad
g^{+ -} \sim \frac{y^2}{2 \rho_+^2}\,,\nonumber\\
&& g^{+ y} \sim \frac{(r_+ - r_-)}{8 M a^2 \rho_+^2 \sin^2\theta}\left(\Sigma+ 2 N a \cos\theta \right)w^+ \, y\,,\\
&& g^{- y} \sim -\left[\frac{(r_+ +M)} { 4 M \rho_+^2}-\frac{(4\pi M T_L-  \Xi)(4 \pi M (T_L+T_R)- \Xi)  }{8 M^2 \rho_+^2 \sin^2\theta}\right]\,  w^{-} \, y\,.\nonumber
\end{eqnarray}

\subsection*{Covariant Charge}

In the previous subsection a set of conformal coordinates were proposed. These coordinates are well-adapted in that they make the deformed quotients of AdS$_3$ on the black hole bifurcation surface geometry explicit for Kerr-Taub-NUT black holes. Similar results were reported for Kerr and Ker--Newman black holes \cite{Haco:2018ske,Haco:2019ggi}. The main feature to notice is that at leading order on the bifurcation surface the metric (\ref{eq:metricBif}) is invariant under the vector fields
\bea
\zeta_{0}=w^+ \partial_+ +\frac{1}{2}\,  y \, \partial_y \,,\qquad \bar{\zeta}_{0}=  w^- \partial_- +\frac{1}{2} \,y\, \partial_y 
\eea
We can consider more general conformal vector fields
\bea\label{eq:vectorfields}
\zeta_{n}= \epsilon_n \partial_+ +\frac{1}{2} \partial_+ \epsilon_n y \partial_y \,,\qquad
\bar{\zeta}_{n}=  \bar\epsilon_n \partial_- +\frac{1}{2} \partial_- \bar\epsilon_n y \partial_y \,.
\eea
and restrict the full set of functions $(\epsilon ,\bar\epsilon)$ so that $(\zeta, \bar\zeta)$ are invariant under $2\pi$ azimuthal rotations is
\bea
\epsilon_n=  2\pi T_R(w^+)^{1+\frac{i n}{2\pi T_R}}\,, \qquad 
\bar\epsilon_n  = 2\pi T_L(w^-)^{1+\frac{i n}{2\pi T_L}}\,.
\eea
Taking $\zeta_n\equiv \zeta(\epsilon_{n})$ and  $\bar\zeta_n=\bar\zeta(\epsilon_{n})$, one can easily verify that the vector fields (\ref{eq:vectorfields}) obey the Lie bracket algebra
\bea
[ \zeta_m,\zeta_n ] = i (n-m)\zeta_{m+n} \,, \qquad
[ \bar\zeta_m, \bar \zeta_n ] =  i (n-m) \bar\zeta_{m+n}\,.
\eea
and the two set commuting with another
\bea
[\zeta_m,\bar\zeta_n]=0\,.
\eea
The generator of a diffeomorphism $\zeta$ is a conserved charge $\mathcal{Q}_\zeta$. Under Dirac brackets, the charges obey the same algebra as the symmetries themselves, up to a possible central term $K_{m,n}$. For well-defined integrable charges one has for the the Dirac bracket algebra
\bea
\{\mathcal{Q}_n,\mathcal{Q}_m\}=(m-n )\mathcal{Q}_{m+n}+ K_{m,n} \, .
\eea
where the central term is given by
\bea
K_{m,n} = \delta_m\mathcal{Q}(\zeta_n,\mathcal{L}_{\zeta_n} g, g)\,.
\eea
 The interpretation of $\delta_m \mathcal{Q}$ is the infinitesimal charge differences between neighboring geometries $g_{\mu\nu}$ and $g_{\mu\nu} + h_{\mu\nu}$ with the variation explicitly given by $h_{\mu\nu}=\mathcal{L}_{\zeta_m} g_{\mu\nu}$ and $h=h^{ab}g_{ab}$. Moreover, under certain conditions, it has been proven that the central term must be constant on the phase space and given by
\bea\label{Kmn}
K_{m,n} = \frac{c_R m^3}{12} \delta_{m+n,0}\,.
\eea
for some constant central charge $c_R$. The last step in this section involves working out the central charge $c_R$ by computing the central term in the Virasoro charge algebra $\delta \mathcal{Q}$.

The general form for the linearized charge associated to a diffeo $\zeta$ on a surface $\Sigma$ with boundary $\partial \Sigma$ is
\bea\label{linearcharge}
\delta_m \mathcal{Q} = \delta \mathcal{Q}_{IW}+\delta \mathcal{Q}_{ct}
\eea
This soft hair is realized as finite covariant right(left)- moving Iyer-Wald Virasoro charges $\delta \mathcal{Q}_{IW}$ and the central terms. To guarantee the integrability and associativity of the charges we consider the linearized charge $\delta \mathcal{Q}_{ct}$ contribution defined in (\ref{eq:counterterm}) from a generalized version of the counterterm valid for the Kerr-Taub-NUT family of black holes in 4D. This term reduces exactly to the Wald-Zoupas counterterm for Kerr black holes in \cite{Haco:2018ske,Haco:2019ggi}, and more importantly, the same counterterm proposed for AdS-Kerr black holes in \cite{Perry:2020ndy}. 

On the one hand, the Iyer-Wald linear charge contribution yields
\bea
\delta \mathcal{Q}_{IW}&=&\frac{1}{16\pi} \int_{\partial\Sigma} * F_{IW}\nonumber \\
&=&\frac{1}{16\pi} \int d\theta dw^+\  (-4\epsilon_{y+\theta- } h^{y-} \bar\zeta^y \Gamma^{-}_{y -})\,.
\eea
And, on the other hand, we find a linear charge contribution from the counterterm
\bea
\delta \mathcal{Q}_{ct} =\frac{1}{16\pi}\int_{\partial\Sigma} {F_{(ct)}}_{ab}\, d \Sigma^{ab}\,
\eea
where $N$ is the volume two-form on the normal bundle to the $\Sigma_{bif}$. 
\bea
{F_{(ct)}}_{ab}=- 4 {N_{d}}^c\nabla_c(\zeta_{[a} {h^d}_{b]}) \,,
\eea
Note that the addition of this counterterm is justified to achieve integrability. The nonzero contributions to $K_{n,m}$ come only from 
\bea\label{charge}
\delta\mathcal{Q}=\frac{1}{16 \pi} \int d\theta dw^+ \epsilon_{\theta + - y} ({F_{(IW)}}^{-y}+{F_{(ct)}}^{-y})
\eea
We find that the integrand
\bea\label{fiw}
{F_{(IW)}}^{-y}=4 h^{y-}\zeta^y\Gamma^-_{y-}\,,
\eea
 and considering ${N_+}^+=1$, ${N_-}^-=-1$ then
 \bea
 {F_{(ct)}}^{-y}&=&-2 \nabla_+(\zeta^- h ^{+y})+2 \nabla_+(\zeta^y h ^{+-})+2 \nabla_-(\zeta^- h ^{-y})-2 \nabla_-(\zeta^y h ^{--})\\
 &=&2 \zeta^y(\nabla_+ h^{+-})+2 (\nabla_-\zeta^-) h^{-y}-2\zeta^y (\nabla_- h^{--})\\
 &=&2 \zeta^y h^{-y}(\Gamma^+_{+y}+\Gamma^-_{-y}-2 \Gamma^-_{-y})\\
 &=& 2 \zeta^y h^{-y} (\Gamma^+_{+y}-\Gamma^-_{-y}) \label{fct}
 \eea 
Note that
\bea
h^{+-}=0\,, \qquad h^{--}=0\,,\qquad  \nabla_+  \zeta^-=0\,,\qquad \nabla_-  \zeta^-=\Gamma^-_{-y}\zeta^y\,,
\eea
and also
\bea
\nabla_+h^{+y}=0\,,\qquad \nabla_+h^{+-}=\Gamma^+_{+y} h^{-y}\,,\qquad \nabla_-h^{-y}=0\,,\qquad \nabla_-h^{--}=2\Gamma^-_{-y} h^{-y}\,.
\eea
Adding (\ref{fiw}) and (\ref{fct}) the terms together in (\ref{charge}) one finds
\bea
\delta\mathcal{Q}&=&\frac{1}{16\pi}\int d\theta dw^+  \epsilon_{+- y \theta} (4 h^{y-}\zeta^y\Gamma^-_{y-}+ 2 \zeta^y h^{-y} (\Gamma^+_{+y}-\Gamma^-_{-y}))\,,\\
&=&\frac{1}{16\pi}\int d\theta dw^+ \frac{8M a \sin\theta  \rho_+^2 }{ y^3}  2 h^{y-}\zeta^y(\Gamma^-_{y-}+\Gamma^+_{+y})\,,
\eea
By working at small $w^+$ and taking the $w^+ \rightarrow 0$ limit (which amounts to approaching $\Sigma_{bif}$ along the future horizon) one finds
\bea 
&& h^{-y}=g^{+-}\partial_+ \zeta^y = \frac{y^3\tilde{\epsilon}''}{4\rho^2_+} \qquad \text{with} \qquad  '=\partial_+\,,\\
&& \int d\theta \sin\theta =2, \,\, \text{and} \qquad \Gamma^+_{+y}+\Gamma^-_{-y}=-\frac{2}{y}.
\eea
Choosing $\zeta$ to be $\zeta_n$ and $\tilde\zeta$ to be $\zeta_m$, the variation becomes
\bea
K_{m,n} = \delta\mathcal{Q}&=&\frac{1}{16\pi}\int d\theta dw^+ \frac{8M a \sin\theta  \rho_+^2 }{ y^3}  2 \left(\frac{y^3\epsilon_m''}{4\rho^2_+}\right) \left(\frac{1}{2} y \epsilon_n' \right)\left(\frac{-2}{y}\right) \,\\
&=&-\frac{1}{16\pi}\int d\theta dw^+ 4M a \sin\theta     {\epsilon_m''} \epsilon_n'  \\
&=&-\frac{1}{16\pi}\int  \frac{dw^+ }{w^+ } 8M a    {\epsilon_m''} \epsilon_n' \,,\\
&=&-\frac{1}{16\pi}(4\pi^2 T_R) 8M a     \frac{i m^3}{2\pi T_R} \delta_{m+n,0}\,\\
&=& - M a  \,  i m^3 \delta_{m+n,0}\,.
\eea
Here we have computed the Dirac bracket of two charges. Passing to the commutator rule of Dirac brackets to commutators $\{.,.\} \rightarrow -i [.,.]$ as introduces a factor of $-i$. The central charge of Kerr-Taub-NUT black holes can be easily identified via (\ref{Kmn})
\bea\label{eq:cr}
c_R= 12 \,a \,M = 12 \, J_0\,.
\eea
The central charge computed from soft hair arguments is in agreement with (\ref{centralcharge}). Recall that the latter computation was totally independent to the one in this section, involving the monodromies of the scalar wave solutions of Klein-Gordon equation. We also note that (\ref{eq:cr}) is a function only of the Kerr-Taub-NUT black hole angular momenta $J_0$. The two Misner strings considered here are attached to the horizon of the black hole, symmetrically distributed, and spinning with $J_s$. See Section \ref{sec:KTN} for more details.  Moreover, the central charge $c_R$ will continue to be independent on the details of the Misner string distribution in these solutions as we show in Section \ref{sec:Misner}.

\section{First Law of Thermodynamics}
\label{sec:FirstLaw}


As given in \cite{Haco:2018ske}, another way of writing the relation between the left, right frequencies and uniquely fixing the identification of the temperatures $(T_L,T_R)$ arises from imposing the first law of thermodynamics
\bea
\delta M=T \delta S+ \Omega \, \delta J\,,
\eea
with the identification
\bea
\omega=\delta M \,,\qquad m= \delta J \,.
\eea

For the Kerr-Taub-NUT black holes we consider the first law (\ref{first law}) for fixed values of the Misner charges, that yields
\bea
\delta M=\pm T_{\pm}\delta S_{\pm}+\Omega_{\pm} \delta J\,.
\eea
and argue that the remarkable validity of the relation 
\bea \label{entropy}
\delta S_{\pm}=\frac{\delta E_L}{T_L}\pm\frac{\delta E_R}{T_R}\,.
\eea
where 
\bea
\delta E_L&=& 2 M (2\pi T_L) \, \delta M \\
\delta E_R&=&2 M (2\pi T_R) \, \delta M -\delta J\,,
\eea
with the right and left temperatures $T_{L,R}$ defined in (\ref{temperatures}). This gives further evidence to the choice of counterterm in the computation of the covariant charges for Kerr-NUT black holes. Any other identifications for the $T_{L,R}$ temperatures will not satisfy the thermodynamic relations (\ref{entropy}).

\section{Kerr-Taub-NUT and general distribution of Misner string}
\label{sec:Misner}

The rotating Taub-NUT black hole solution can also include a dimension-full physical parameter which governs the overall distribution and strength of the Misner string. The parameter $s$ parametrizing the Misner distribution can be formally added by performing in the black hole metric (\ref{KerrTaubNUT}) the `large coordinate transformation'
\bea\label{eq:largeccord}
t\rightarrow t + 2 s \phi\,.
\eea
The absence of closed timelike and null geodesics requires $|s/N|\le 1$. In particular, when $s=+N$, there is only one Misner string and it is located on the north pole $(\cos \theta =+1)$ axis, while the south pole $(\cos \theta=-1)$ axis is completely regular. The choice $s=-N$ corresponds to the opposite situation, while for $s= 0$ the two Misner strings are ‘symmetrically distributed’ and both axes are ‘equally singular’. See \cite{Bordo:2019rhu} for details.

The solution takes the form (\ref{KerrTaubNUT}) with the following definitions of the functions
\bea
&&\rho^2=r^2+(N+a \cos\theta)^2\,,\,\,\, \Delta=r^2-2 M r-N^2+a^2\,,\\
&& \Xi=a\sin^2\theta - 2N \cos\theta- 2 s\, ,\,\,\,\,\,\, \Sigma=r^2+N^2+a^2 -2 a s\,
\eea
It is important to note that the location of the outer/inner horizons remain invariant under the change of the Misner string distribution. These are defined via $\Delta=0$ are $r_{\pm}=M\pm\sqrt{M^2-a^2+N^2}$.

The black hole horizon can be assigned the following entropy
\bea\label{eq:entropyS}
S_{\pm}=\pi \, (r_{\pm}^2+a^2+N^2- 2 a s)\,.
\eea
Note that this quantity explicitly depends on the parameter $s$. This means that the total Bekenstein-Hawking entropy actually
differs from the special case where the strings are symmetrically distributed with the $s=0$ case above. This suggests that the identification of the dual CFT quantities will incorporate the dependence of the Misner string distribution parameter $s$.

With the explicit expression for the black hole entropy in hand (\ref{eq:entropyS}), let us now turn towards the monodromies (\ref{monodromies}). Following the same steps as in our previous sections we identify the conjugate variables 
\bea\label{eq:conformalcoord}
t_L=2\pi T_L \,\phi- \frac{1}{2M}\,t,\qquad
t_R= 2\pi T_R \, \phi\,,
\eea
with $(t,\phi)$ the Boyer-Lindquist coordinates and
\bea
T_R=\frac{r_+-r_-}{4\pi a}\,,\qquad T_L=\frac{M^2+N^2 - a s}{2\pi a M}=\frac{(r_++r_-)^2+4 N^2- 4 a s}{4\pi a (r_++r_-)}\,.
\eea
Employing Cardy formula 
\bea
S_{\pm}= S_{Cardy}=\frac{\pi^2}{3} c \, (T_L\pm T_R)\,,
\eea
one finds that the central charge $c=c_L=c_R$ is defined by
\bea
c=12 a M\,.
\eea
The same central charge is found from the soft hair arguments.
Note that the large diffeomorphism (\ref{eq:largeccord}) in addition to the coordinate transformation (\ref{eq:conformalcoord}) will lead to a metric containing a deformed quotient of $AdS_3$ close to the birfurcation (\ref{eq:metricBif}). While we do not present here the full derivation of the soft hair charges, we argue that the structure of the metric (\ref{eq:metricBif}) close to the bifurcation surface suffices to straightforwardly carry out the steps in Section \ref{sec:ConfCoord}.  

Having now applied the soft hair holds for rotating black holes with charges, we now turn to general classes of 4D black holes. We will argue that the generalized counterterm (\ref{eq:counterterm}) leads to well defined soft hair linearized charges.

\section{Soft hair charges on general classes of 4D black holes}
\label{sec:AllBH}

The complete family of (non-accelerating) electric and magnetically charged black-hole like space-times in four space-time dimensions \cite{Griffiths:2005qp} can be written in Boyer-Lindquist coordinates \footnote{Note that the metric here differs respect to  \cite{Griffiths:2005qp}  by a coordinate transformation $\phi^{(there)} \rightarrow  \phi/ \Xi$ and $t^{(there)} \rightarrow t+ 2 N \, \phi / K$} as
\bea \label{allBH}
ds^2&=&-\frac{Q}{\rho^2}\left[dt+(2N \cos\theta- a\sin^2\theta)\,\frac{d\phi}{K}\right]^2+\frac{\rho^2}{Q}dr^2 \\
&& +\frac{{P}}{\rho^2}\left[a\,dt-(r^2+a^2+N^2)\,\frac{d\phi}{K}\right]^2+\frac{\rho^2}{{P}}\sin^2\theta d\theta^2 \,,
\eea
where
\bea
\rho^2&=&r^2+(N+a\cos\theta)^2\\
P&=&\sin^2\theta \, \left(1+  \frac{\Lambda}{3}  \, 4 a N \cos\theta+ \frac{\Lambda}{3} \,a^2 \cos^2\theta \right)\\
Q&=&(\omega^2 k+e^2+g^2)-2\,m \,r+\epsilon r^2-\frac{\Lambda}{3} \,r^4\,,\\
K&=&1-a^2/L^2\,,
\eea
with $\epsilon, n$ and $k$ are given by
\bea
\epsilon&=& \frac{\omega^2 k}{a^2-N^2} - (a^2+3 N^2) \, \frac{\Lambda}{3}\,,\\
n&=& \frac{\omega^2 k N}{a^2-N^2}+ (a^2- N^2)\,N \, \frac{\Lambda}{3}\,,\\
k&=& \left(1- N^2 \Lambda\right)\left(\frac{\omega^2}{a^2-N^2}\right)^{-1}\,.
\eea
satisfying Einstein's equations 
\bea
&&R_{\mu\nu} -(1/2) \, R\, g_{\mu\nu} + \Lambda \, g_{\mu\nu}=2 \left(F_{\mu\alpha}{F_{\nu}}^{\alpha} -(1/4)\,g_{\mu\nu}F_{\alpha\beta}F^{\alpha\beta}\right)\,,\\
&&\partial_{\mu} \left(\sqrt{-g} \, F^{\mu\nu}\right)=0\,,
\eea
 where $\Lambda=-3/L^2$ is the cosmological constant, and the Faraday tensor $F_{\mu\nu}=\partial_{\mu}\tilde{A_{\nu}}-\partial_{\nu}\tilde{A_{\mu}}$ is given in terms of the electromagnetic potential 
 \bea
  \tilde{A}=- \frac{e \, r }{\rho^2}\,\left[ \,dt-  \,\Xi\,\,\frac{d\phi}{K} \right]+ \frac{g}{\rho^2} \, \left[p\,dt - (\rho^2 \cos\theta+p \,\Xi)\,\,\frac{d\phi}{K}\right]\,.
  \eea
  where
   \bea \label{pequation}
  \Xi =a \sin^2\theta- 2 N \cos \theta \,,\qquad p=N+a\cos\theta\,.
  \eea
  This vector potential reduces to the gauge field expression in \cite{Podolsky:2006px} when the NUT charge is set to zero $N=0$.

This solution to Einstein's equations contains seven parameters: $m, e, g, a , N, \Lambda $ and $\omega$ which represent respectively the mass, electric charge, magnetic charge, rotation,  nut charge, cosmological constant and remaining scaling freedom.  Of these, the first six can be varied independently, and $\omega$ can be set to any convenient value if $a$ or $N$ are not both zero. The event horizon of the black hole is located at $r=r_+$ where $Q(r_+)=0$. The metric (\ref{allBH}) has also a Cauchy horizon $r=r_-$ when $Q(r_-)=0$ and $0<r_- <r_+$.

To elucidate the thermodynamics of the metric, we now consider the black hole event horizon area computed from (\ref{allBH}) by integrating
\bea
Area \equiv \int \sqrt{g_{\phi\phi} \, g_{\theta\theta}} \, d\theta \,d\phi=\frac{4\pi\,(r_{\pm}^2+a^2+N^2)}{ K }\,.
\eea
such that the entropy yields
\bea
S_{\pm}\equiv \frac{Area}{4}=\frac{\pi\,(r_{\pm}^2+a^2+N^2)}{ K }\,.
\eea
The expressions for the Hawking temperature and angular velocity are similar to those for the uncharged AdS-Kerr-Taub-NUT black hole presented in \cite{Rodriguez:2021hks}
\bea
T_{\pm}= \frac{Q'_+}{4\pi(r_{\pm}^2+a^2+N^2)} \,,\qquad \Omega_{\pm}=\frac{a\, K}{r_{\pm}^2+a^2+N^2}
\eea
However, in present case, the outer and inner horizons $r_{\pm}$ involve the electric and magnetic charges $e$ and $g$. One can see from these thermodynamic expressions that in the vanishing NUT charge ($N=0$) limit we recover the quantities given in \cite{Podolsky:2006px}. Further, these quantities are also consistent with \cite{Rodriguez:2021hks} for $e=g=0$. 

As advocated in \cite{Haco:2018ske,Haco:2019ggi,Haco:2019paj}, the null-tetrad representation can be used in the derivation of the near black hole bifurcation surface region to simplify the computations. In particular, the corresponding null-tetrad representation for the  space-time (\ref{allBH}) is
\bea
g^{ab}= - l^a n^b-n^a l^b+m^a \bar{m}^b+\bar{m}^a m^b
\eea
with $a,b=t,r,\theta,\phi$ and the null tetrad defined as
\bea
l^a&=&\omega \left[(r^2+a^2+N^2) \,\partial_t-Q  \,\partial_r +a \,K\,\partial_{\phi}\right]\\
n^a&=&\frac{\omega }{ {2 \, Q\, (r^2+ p^2)}}   \left[(r^2+a^2+N^2) \, \partial_t+ Q \, \partial_r+a \, K\, \partial_{\phi}\right]\\
m^a&=&-\frac{\omega }{ \sqrt{2\, P\, (r^2+p^2)}}  \left[\frac{ p^2-(a^2+N^2)}{a}\, \partial_t+ i \frac{\,P }{\sin\theta}\,\partial_{\theta}- K\, \partial_{\phi}\right]\,.
\eea
and $p$ defined in (\ref{pequation}).

One may now wonder if all 4D black holes (\ref{allBH}) exhibit a conformal structure in the near bifurcation surface geometry.
As we will see in the next section, it is in fact that adapted $(w_{\pm},y,\theta)$ conformal coordinates which naturally arise from
the study of the Kerr black holes give rise to this universal near horizon geometry feature in black holes. The conformal coordinates were useful in the context of extending the Kerr/CFT proposal to non-extremal black holes \cite{Castro:2010fd} by explicitly describing  the conformal symmetry of the space on which the field propagates via e.g. the Klein-Gordon equation rather than the symmetry of the spacetime geometry. In the next section of this paper, however, we will give a different interpretation. In particular, we will connect the close to the bifurcation surface geometry  to the soft hair analysis and CFT dual.

\subsection{Near bifurcation surface metric for generic black holes}

In order to better understand these local conformal symmetries in the metrics and the validity of the counterterms in soft-hair approach leading to the Hawking entropy we will explicitly compute the near bifurcation surface metrics for all generic black holes solutions of Einstein equations in 4D \cite{Plebanski:1976gy}. We will argue that stationary non-accelerating black hole metrics of the Plebański–Demiański class (\ref{allBH}), to leading order at the bifurcation surface completely determine the soft hair charge and its Bekenstein-Hawking entropy.

We start by considering the perturbative expansion of a metric around the event horizon $r=r_+$. The event horizon radius $r_+$ sets the size of the black hole (\ref{allBH}). We define the zoom in coordinates on the event horizon as
\bea
r&\rightarrow &r_+ + \frac{w^+ w^-}{y^2} \, Q'(r_+)\, ,\\
\phi &\rightarrow & (2 \alpha)^{-1} \log\left( \frac{w^+ y^2}{w^-}\right)\,,\\
t &\rightarrow & (2\delta)^{-1} \log\left(\frac{w^- y^2}{w^+}\right) - \frac{\gamma}{\delta} \phi\,.
\eea
By applying an expansion for $w^+<<r_+$, the near  bifurcation surface metric can be cast into an elegant simple geometry, defined by
\bea\label{near horizon metric}
ds^2 \approx\frac{1}{y^2} \left(4 \, \rho_+^2 dw_+ \, dw_- +\frac{4 \,a^2 P}{\delta^2 \rho_+^2} dy^2 \right)+ \frac{\rho_+^2 \, \sin^2\theta}{P} d\theta^2
\eea
for the following choices of the parameters\\
\bea
\alpha=\frac{Q'_+}{2 \,a \,K} \, ,\qquad\,  \gamma= \frac{(\Omega_+ +\Omega_- )}{(\Omega_--\Omega_-)}\,\alpha\,,\qquad \delta=-\Omega_+ (\gamma+\alpha)\,.
\eea
and where for any function $X$ we have defined its evaluation on the horizon $X_+=X(r_+)$. The universal expression for the near bifurcation surface metric of black holes that we find signals at an underlying connection to a locally defined $SL(2,R)_L \times SL(2,R)_R$ symmetry of the near-horizon geometry and near-region scalar field equation. For more details on (\ref{near horizon metric}), we refer the interested reader to \cite{Rodriguez:2021hks} where symmetries and its thermodynamics is studied.


\subsection{Central charge}

The immediate goal of this subsection is to identify the right and left-temperatures arising from the CFT for general classes of 4D black holes. The systematic approach presented in the previous section unveiled that 4D black holes mirror the geometry of the 3D BTZ black holes in the regions close to the bifurcation surface. The periodicities analysis under the azimuthal identification $\phi\rightarrow \phi+2\pi$  of the conformal variables yields
\bea
w^+\sim e^{2\pi \alpha}\,,\qquad w^-\sim e^{2\pi \gamma}\,,\qquad y\sim e^{\pi (\alpha+\gamma)}
\eea
This is the same as the identification employed in \cite{Maldacena:1998bw} that turns AdS$_3$ in Poincare coordinates into BTZ with temperatures 
\bea
\alpha\rightarrow 2\pi T_R\,, \qquad \gamma\rightarrow 2\pi T_L\,.
\eea
Under the holographic duality the dimensionless left and right temperatures are then
\bea\label{temps}
T_R=\frac{Q'}{4\pi a\, K}\,,\qquad T_L=\frac{(\Omega_+ +\Omega_- )}{(\Omega_--\Omega_-)}\,\frac{Q'}{4\pi a\, K}\,.
\eea
According to the Cardy formula the entropy for a unitary CFT obeys
\bea
S_{\pm}= S_{Cardy}=\frac{\pi^2}{3} c \, (T_L\pm T_R)\,,
\eea
Using (\ref{temps}) we find the microscopic entropy for the dual black hole (\ref{allBH}) regarded the central charge of the CFT $c=c_L=c_R$ is defined by
\bea\label{centralchargy}
c=-\frac{6 a }{\delta}\,.
\eea
While this very general analysis gives the central charge of the dual CFT, we need to construct the linearized charge associated to a diffeo (\ref{linearcharge}) acting on the horizon and see if they are finite.
 It can easily be verified that the general form of the near bifurcation surface metric (\ref{near horizon metric}) gives a finite linearized charge yielding (\ref{centralchargy})

\section{Discussion}

In this work we have constructed the 2D CFT identifications for the 4D Kerr-Taub-NUT and Plebianski-Demianski 
spacetimes. Extending three different approaches  -- monodromies, soft-hair and thermodynamics -- for these 4D black hole geometries we have found explicit expressions for the dual temperatures $(T_L,T_R)$ and central charge. Remarkably, comparing the microscopic entropy with the the macroscopic area law, we see an impressive and detailed match for general $(m, a , N, g, e, \Lambda)$ black holes. We expect this to have implications for the hypothesis that hidden conformal symmetry explains the leading microstate degeneracy for such 4D black hole backgrounds.


We conclude with two miscellaneous observations.

As we stated earlier, the choice of the boundary counterterm relevant for the linearized soft-hair charge contribution is not unique. Originally the Wald-Zoupas counterterm was proposed to reproduce the area law for the Kerr(Newman) black hole \cite{Haco:2018ske,Haco:2019ggi}. In \cite{Perry:2020ndy}, the authors found that for AdS-Kerr black holes a more general proposal (\ref{eq:counterterm}) was expected for constructing the 2D CFT that was consistent with all other results from the monodromy techniques, the thermodynamic analysis and all extremal black holes studies (Table \ref{Table1}). Extending the analysis of \cite{Perry:2020ndy} for Kerr-Taub-NUT and the complete family of 4D black holes we explicitly showed that the more general counterterm (\ref{eq:counterterm}) 
gives rise to well-defined charge and gives a central charge $c_R = c_L$.  Our argument is sharpened version of those previously made and is perhaps the most
general in spirit given that it works for the complete family of 4D black holes.

We do not herein proved uniqueness of the counterterm, but instead presented extensive evidence of the applicability of a more general counterterm for the general classes of 4D black holes. This indicates that the applicability of the Wald-Zoupas is indeed limited. It is also worth emphasizing that the counterterm (\ref{eq:counterterm}) reduces to the Wald-Zoupas counterterm for Kerr-Newman black holes, and hence includes the results found in \cite{Haco:2018ske,Haco:2019ggi}. The observation of the inconsistent results derived from the extension of the Wald-Zoupas counterterm to Kerr-Taub-NUT can be found in Appendix \ref{AppendixA}. While we kept the identification of the dual temperatures (\ref{temperatures}) in the computations, we found that the charge associated with the diffeormorphisms on the horizon supported with the Wald-Zoupas counterterm led to a central charge defined by $c_L\ne c_R$. Moreover, the central charge compatible with the Cardy formula was further not quantized -- in contrast to our results where $c_L=c_R= 12 J_0$ -- and is inconsistent with previously know results from extensions of the Kerr/CFT conjecture for extreme black holes. 

Our final comment in this concluding section will focus on the Kleinian aspects of the (Lorentzian) Kerr-Taub-NUT black holes and 2D CFTs. For the sake of simplicity, let us consider the Kerr-Taub-NUT metric, which is determined by three parameters $(M, N, a)$. Kleinian signature metric \cite{Barrett:1993yn} as shown in \cite{Crawley:2021auj,Adamo:2021lrv,Geyer:2020iwz}  for black holes, can construct its Kleininan signature metric following the analytic continuation $t\rightarrow i \, t$ and $\rightarrow i \, \theta$ supplemented by $a\rightarrow i \, a$ and $N\rightarrow i N$ required by the reality of the metric. When the mass and NUT charge are equal, $M = \pm N$ the geometry is self-dual for any value of the Kerr rotation parameter a. In fact, a can be absorbed into a shift of the  $r$-coordinate. Via a coordinate transformation involving both $(r, \theta)$ the geometry can be shown to be diffeomorphic to the self-dual Kleinian Taub-NUT metric \cite{Crawley:2021auj}. Inspecting the 2D CFT quantities we derived for the Kerr-Taub-NUT metrics, we argue that the Kleinian Kerr-Taub-NUT metric will have (if any) a 2D CFT dual with temperatures
  \bea
  T_L= 0\,,\qquad T_R=\frac{1}{2\pi}
  \eea
and central charge
\bea
c= 12 J\,.
\eea
Remarkably, this is exactly the opposite representations of the one copy of the Virasoro algebra found for the extreme Kerr black holes within the Kerr/CFT correspondence. 

Kleinian metrics have a natural interpretation in terms of scattering amplitudes. Perhaps interesting simplifications and structures occur in a 2D CFT description of these spacetimes, which can lead to further implications for constructing scattering amplitudes. 

There are still plenty of open questions. Is there a holographic dual for the Kleinian Kerr-Taub-NUT? Does an explicit microscopic construction of 4D black holes exist beyond the semiclassical limit considered here? These appear to be interesting avenues for future pursuit.

\section*{Acknowledgement}

The work of MP is supported by an STFC consolidated grant ST/L000415/1, String Theory, Gauge Theory and Duality. MJR is partially supported through the 
 NSF grant PHY-2012036, RYC-2016-21159, CEX2020-001007-S and PGC2018-095976-B-C21, funded by MCIN/AEI/10.13039/501100011033.
 

\appendix

\section{Alternative counterterm}
\label{AppendixA}

The Wald-Zoupas term is of the form
\bea
\delta \mathcal{Q}_{ct}&=&\frac{1}{16\pi} \int_{\partial\Sigma} \zeta(*X)\nonumber \\
&=&\frac{1}{16\pi} \int d\theta dw^+ \epsilon_{y+\theta- } 2 \, h^{y-} \bar\zeta^y ( n_+  \partial_y l^+ + n_+ \Gamma^+_{y+}l^+)  \nonumber \\
&=&\frac{1}{16\pi} \int d\theta dw^+ \epsilon_{y+\theta- } (-2) h^{y-} \bar\zeta^y \left(\frac{2 T_R}{y(T_L+T_R)}+\Gamma^{+}_{y +}\right)\,,
\eea
In order to have null vectors normal to the bifurcation surface that are periodic we take $l^{\mu}\sim \frac{1}{\sqrt{2} \,  \rho_+}y^{2 \,T_R/(T_L+T_R)}\partial_+$ and $n^{\mu}\sim-\frac{1}{\sqrt{2} \, \rho_+} y^{2 \, T_L/(T_L+T_R)}\partial_-$ normalized such that $l . n=-1$.\\

The charge associated to the diffeomerphism $\zeta_n$ reduces to a boundary integral
\bea
\delta \mathcal{Q} = \frac{1}{16\pi} \int d\theta dw^+ \epsilon_{y+\theta- } (-2) h^{y-} \bar\zeta^y  \left(2\, \Gamma^{-}_{y -}+\frac{2\, T_R}{y(T_L+T_R)}+\Gamma^{+}_{y +}\right)
\eea
One finds after a simple integral the central charge gives
\bea
c_R= \frac{12 M a}{((T_L+T_R)(T_L+T_R-\Theta)}\left[ (T_L+T_R)^2 - T_L \Theta \right]
\eea
where we labeled $\Theta=N^2/(2\pi a M) $. In our case when the NUT parameter vanishes we recover previous results in \cite{Haco:2018ske} for the Kerr black hole.
\subsection*{Christoffel symbols }
The Christoffel symbols relevant for this computation are found to be

\begin{tiny}
\bea
\Gamma^{+}_{+y}&=&\frac{ a^2K(K-4 \pi M (2 T_L+T_R) )-M (r_+-r_-)(\Sigma+2 aN\cos \theta  )-\rho_+^4-2 a M \left(a \sin ^2\theta  (M+r_+)-8 \pi ^2 a M T_L (T_L+T_R)\right)}{y \rho_+^4}\nonumber\\
\Gamma^{-}_{-y}&=&-\frac{ a^2 K (K-4 \pi  M (2 T_L+T_R))-M (r_+-r_-)(\Sigma+2 aN\cos \theta  )+\rho_+^4-2 a M \left(a \sin^2 \theta  (M+r_+)-8 \pi ^2 a MT_L (T_L+T_R)\right)}{y \rho_+^4}\nonumber
\eea
\end{tiny}

where
\bea
\Gamma^+_{+y}+\Gamma^-_{-y}=-\frac{2}{y}.
\eea

And the integrals are
\bea
\int_0^{\pi} d\theta \sin\theta \,  \Gamma_{+y}^+=-\frac{2\, (r_++r_-)}{ y\,r_+}\,,\qquad \int_0^{\pi} d\theta \sin\theta \, \Gamma_{-y}^-=-\frac{2\, (r_+-r_-)}{ y\,r_+}\eea


\begin{thebibliography}{99}

\bibitem{Guica:2008mu}
M.~Guica, T.~Hartman, W.~Song and A.~Strominger,
``The Kerr/CFT Correspondence,''
Phys. Rev. D \textbf{80} (2009), 124008
doi:10.1103/PhysRevD.80.124008
[arXiv:0809.4266 [hep-th]].

\bibitem{Hartman:2008pb}
T.~Hartman, K.~Murata, T.~Nishioka and A.~Strominger,
``CFT Duals for Extreme Black Holes,''
JHEP \textbf{04} (2009), 019
doi:10.1088/1126-6708/2009/04/019
[arXiv:0811.4393 [hep-th]].

\bibitem{Castro:2010fd}
A.~Castro, A.~Maloney and A.~Strominger,
``Hidden Conformal Symmetry of the Kerr Black Hole,''
Phys. Rev. D \textbf{82} (2010), 024008
doi:10.1103/PhysRevD.82.024008
[arXiv:1004.0996 [hep-th]].

\bibitem{Haco:2018ske}
S.~Haco, S.~W.~Hawking, M.~J.~Perry and A.~Strominger,
``Black Hole Entropy and Soft Hair,''
JHEP \textbf{12} (2018), 098
doi:10.1007/JHEP12(2018)098
[arXiv:1810.01847 [hep-th]].


\bibitem{Haco:2019ggi}
S.~Haco, M.~J.~Perry and A.~Strominger,
``Kerr-Newman Black Hole Entropy and Soft Hair,''
[arXiv:1902.02247 [hep-th]].


\bibitem{Perry:2020ndy}
M.~Perry and M.~J.~Rodriguez,
``Central Charges for AdS Black Holes,''
[arXiv:2007.03709 [hep-th]].

\bibitem{Lu:2008jk}
H.~Lu, J.~Mei and C.~N.~Pope,
``Kerr/CFT Correspondence in Diverse Dimensions,''
JHEP \textbf{04} (2009), 054
doi:10.1088/1126-6708/2009/04/054
[arXiv:0811.2225 [hep-th]].

\bibitem{Chen:2020nyh}
L.~Q.~Chen, W.~Z.~Chua, S.~Liu, A.~J.~Speranza and B.~d.~ .~Torres,
``Virasoro hair and entropy for axisymmetric Killing horizons,''
Phys. Rev. Lett. \textbf{125} (2020), 241302
doi:10.1103/PhysRevLett.125.241302
[arXiv:2006.02430 [hep-th]].


\bibitem{Chandrasekaran:2020wwn}
V.~Chandrasekaran and A.~J.~Speranza,
``Anomalies in gravitational charge algebras of null boundaries and black hole entropy,''
JHEP \textbf{01} (2021), 137
doi:10.1007/JHEP01(2021)137
[arXiv:2009.10739 [hep-th]].


\bibitem{Castro:2013kea}
A.~Castro, J.~M.~Lapan, A.~Maloney and M.~J.~Rodriguez,
``Black Hole Monodromy and Conformal Field Theory,''
Phys. Rev. D \textbf{88} (2013), 044003
doi:10.1103/PhysRevD.88.044003
[arXiv:1303.0759 [hep-th]].

\bibitem{Wang:2010qv}
Y.~Q.~Wang and Y.~X.~Liu,
``Hidden Conformal Symmetry of the Kerr-Newman Black Hole,''
JHEP \textbf{08} (2010), 087
doi:10.1007/JHEP08(2010)087
[arXiv:1004.4661 [hep-th]].


\bibitem{Chen:2010xu}
B.~Chen and J.~Long,
``Real-time Correlators and Hidden Conformal Symmetry in Kerr/CFT Correspondence,''
JHEP \textbf{06} (2010), 018
doi:10.1007/JHEP06(2010)018
[arXiv:1004.5039 [hep-th]].

\bibitem{Setare:2010cj}
M.~R.~Setare and V.~Kamali,
``Hidden Conformal Symmetry of Extremal Kerr-Bolt Spacetimes,''
JHEP \textbf{10} (2010), 074
doi:10.1007/JHEP10(2010)074
[arXiv:1011.0809 [hep-th]].

\bibitem{Rodriguez:2021hks}
N.~H.~Rodr\'\i{}guez and M.~J.~Rodriguez,
``First Law for Kerr Taub-NUT AdS Black Holes,''
[arXiv:2112.00780 [hep-th]].

\bibitem{Crawley:2021auj}
E.~Crawley, A.~Guevara, N.~Miller and A.~Strominger,
``Black Holes in Klein Space,''
[arXiv:2112.03954 [hep-th]].

\bibitem{Bordo:2019rhu}
A.~Ballon Bordo, F.~Gray, R.~A.~Hennigar and D.~Kubiz\v{n}\'ak,
``The First Law for Rotating NUTs,''
Phys. Lett. B \textbf{798} (2019), 134972
doi:10.1016/j.physletb.2019.134972
[arXiv:1905.06350 [hep-th]].

\bibitem{Chanson:2020hly}
A.~B.~Chanson, J.~Ciafre and M.~J.~Rodriguez,
``Emergent black hole thermodynamics from monodromy,''
Phys. Rev. D \textbf{104} (2021) no.2, 024055
doi:10.1103/PhysRevD.104.024055
[arXiv:2004.14405 [gr-qc]].

\bibitem{Puletti:2021msz}
V.~G.~M.~Puletti and V.~L.~Martin,
``Hidden Conformal Symmetry in Higher Derivative Dynamics,''
[arXiv:2112.14544 [hep-th]].

\bibitem{Aggarwal:2019iay}
A.~Aggarwal, A.~Castro and S.~Detournay,
``Warped Symmetries of the Kerr Black Hole,''
JHEP \textbf{01} (2020), 016
doi:10.1007/JHEP01(2020)016
[arXiv:1909.03137 [hep-th]].

\bibitem{Novaes:2014lha}
F.~Novaes and B.~Carneiro da Cunha,
``Isomonodromy, Painlev\'e transcendents and scattering off of black holes,''
JHEP \textbf{07} (2014), 132
doi:10.1007/JHEP07(2014)132
[arXiv:1404.5188 [hep-th]].

\bibitem{Maldacena:1998bw}
J.~M.~Maldacena and A.~Strominger,
``AdS(3) black holes and a stringy exclusion principle,''
JHEP \textbf{12} (1998), 005
doi:10.1088/1126-6708/1998/12/005
[arXiv:hep-th/9804085 [hep-th]].

\bibitem{Brown:1986nw}
J.~D.~Brown and M.~Henneaux,
``Central Charges in the Canonical Realization of Asymptotic Symmetries: An Example from Three-Dimensional Gravity,''
Commun. Math. Phys. \textbf{104} (1986), 207-226
doi:10.1007/BF01211590


\bibitem{Griffiths:2005qp}
J.~B.~Griffiths and J.~Podolsky,
``A New look at the Plebanski-Demianski family of solutions,''
Int. J. Mod. Phys. D \textbf{15} (2006), 335-370
doi:10.1142/S0218271806007742
[arXiv:gr-qc/0511091 [gr-qc]].


\bibitem{Podolsky:2006px}
J.~Podolsky and J.~B.~Griffiths,
``Accelerating Kerr-Newman black holes in (anti-)de Sitter space-time,''
Phys. Rev. D \textbf{73} (2006), 044018
doi:10.1103/PhysRevD.73.044018
[arXiv:gr-qc/0601130 [gr-qc]].

\bibitem{Haco:2019paj}
S.~Haco,
``Large Gauge Transformations and Black Hole Entropy,''
doi:10.17863/CAM.48185



\bibitem{Plebanski:1976gy}
J.~F.~Plebanski and M.~Demianski,
``Rotating, charged, and uniformly accelerating mass in general relativity,''
Annals Phys. \textbf{98} (1976), 98-127
doi:10.1016/0003-4916(76)90240-2

\bibitem{Barrett:1993yn}
J.~W.~Barrett, G.~W.~Gibbons, M.~J.~Perry, C.~N.~Pope and P.~Ruback,
``Kleinian geometry and the N=2 superstring,''
Int. J. Mod. Phys. A \textbf{9} (1994), 1457-1494
doi:10.1142/S0217751X94000650
[arXiv:hep-th/9302073 [hep-th]].

\bibitem{Adamo:2021lrv}
T.~Adamo, L.~Mason and A.~Sharma,
``Celestial $w_{1+\infty}$ Symmetries from Twistor Space,''
SIGMA \textbf{18}, 016 (2022)
doi:10.3842/SIGMA.2022.016
[arXiv:2110.06066 [hep-th]].

\bibitem{Geyer:2020iwz}
Y.~Geyer, L.~Mason and D.~Skinner,
``Ambitwistor strings in six and five dimensions,''
JHEP \textbf{08}, 153 (2021)
doi:10.1007/JHEP08(2021)153
[arXiv:2012.15172 [hep-th]].

%
%
%
%
%
%



\end{thebibliography}
\end{document}